# Experimental Realization of a Reflections-Free Compact Delay Line Based on a Photonic Topological Insulator

Kueifu Lai[1], Tzuhsuan Ma[1], Xiao Bo[2,3], Steven Anlage[2,3], and Gennady Shvets[1,*]

[1] Department of Physics, The University of Texas at Austin, Austin, Texas 78712, USA

[2] Center for Nanophysics and Advanced Materials, Department of Physics, University of Maryland, College Park, Maryland 20742-4111, USA

[3] Department of Electrical and Computer Engineering, University of Maryland, College Park, Maryland 20742-3285, USA

*gena@physics.utexas.edu

**Abstract**

Electromagnetic (EM) waves propagating through an inhomogeneous medium inevitably scatter whenever the medium's electromagnetic properties change on the scale of a single wavelength. This fundamental phenomenon constrains how optical structures are designed and interfaced with each other. Recent theoretical work indicates that electromagnetic structures collectively known as photonic topological insulators (PTIs) [1, 2, 3, 4, 5, 6, 7, 8, 9, 10, 11, 12, 13] can be employed to overcome this fundamental limitation [14], thereby paving the way for ultra-compact photonic structures that no longer have to be wavelength-scale smooth. Here we present the first experimental demonstration of a photonic delay line based on topologically protected surface electromagnetic waves (TPSWs) between two PTIs which are the EM counterparts of the quantum spin-Hall topological insulators [15, 16, 17, 18] in condensed matter. Unlike conventional guided EM waves that do not benefit from topological protection, TPSWs are shown to experience multi-wavelength reflection-free time delays when detoured around sharply-curved paths, thus offering a unique paradigm for compact and efficient wave buffers and other devices.

The existence of localized TPSWs at the PTI's edge [6, 7, 10], or at an interface between two PTIs with different electromagnetic properties [1, 3, 11, 12, 14], holds great promise for photonic applications. Their scattering-free propagation along sharply-curved paths opens exciting opportunities across the electromagnetic spectrum, including optical isolators [19, 20], multiple-input multiple-output communications systems [21], and topologically robust broadband optical buffers and time delay lines [22, 23]. Remarkably, while the latter set of applications was the original motivation [6] for PTI development, an experimental demonstration of such functionality has been elusive. For example, in one successful implementation of topologically protected edge transport that utilized an ensemble of high-Q resonators [7], statistical properties of time delays were measured [24]. However, the combination of finite disorder and sharp resonances makes the development of a single-channel delay line in a given photonic structure extremely challenging.

An alternative PTI platform emulates the quantum spin Hall (QSH) [15, 16] effect by introducing a photonic analog of spin-orbital interaction using bianisotropic metamaterials [11], as well as uniaxial [12] or bianisotropic [14] metawaveguides. The spin degree of freedom can then be interpreted as the phase relationship between transverse electric (TE) and magnetic (TM) modes of the metamaterial – in-phase for the spin-up and out-of-phase for the spin-down states. In this Letter, an interface between two QSH PTIs is used to experimentally demonstrate a single-channel topologically protected delay line that employs only the edge modes and is not influenced by the bulk modes. Strong suppression of the bulk modes with respect to the edge modes is crucial because the former are strongly affected by lattice disorder and are subject to localization-based transport [25] that affects the delay time [24].

The specific platform used in this work is shown in Fig. 1a. The QSH PTI is comprised of the parallel-plate metal waveguide sandwiching a periodically arranged hexagonal array of metallic cylinders attached to one of the two metal plates and separated by a finite gap from the opposite plate. The simulated photonic band structure (PBS) of the PTI shown in Fig. 1b (see the caption and Methods for the physical dimensions, and the details of numerical simulations and measurements) reveals a complete topological band gap (gray-shaded area) of the bulk PTI. The bandgap was demonstrated by measuring a 30dB transmission drop in the $\left|f - f_G^{(\text{top})}\right| < \Delta f_{BG}/2$ frequency range (black line in Fig. 1c: $f_G^{(\text{top})} \approx 6.1\text{GHz}$ and $\Delta f_{BG} \approx 0.07 f_G$) when all rods are attached to the top plate. A second, *topologically trivial* bandgap centered at $f_G^{(\text{tri})} \sim 3\text{GHz}$ (not shown), was also observed, and will be discussed later when comparing the topologically protected and trivial guided modes. The structure can be scaled down to operate at telecom frequencies without major performance degradation by either plasmonic effects or high Ohmic loss (see Methods).

The topological spin-Chern index [26] of the electromagnetic modes propagating below the bandgap changes sign [14] when the rods are re-attached from the top to the bottom plate (see SI, section S1). Therefore, the Chern number is reversed across the wall between two QSH PTIs domains ("claddings") with the rods attached to the opposite plates as shown in Fig. 1a. Such topological waveguide is expected [3] to support four TPSWs plotted in Fig. 1b as red lines: two spin-up states propagating in the forward, and two spin-down states propagating in the backward directions (see section S2 of the SI). In the absence of spin-flipping perturbations, backscattering is prohibited for the TPSWs. Their existence across the entire bandgap is experimentally demonstrated (Fig. 1c) by measuring the $\sim 30 dB$ transmission enhancement (red line) over that through the bulk PTI (black line). Therefore, the electromagnetic waves excited by the launching antenna inside the bandgap do not evanescently tunnel through the PTI's bulk. Instead, they couple to the surface mode and propagate unimpeded towards the probe antenna. The spatial localization of the surface mode to a small fraction of the wavelength on either side of the interface is established by mapping the field profile in the $y$-direction (Fig. 2a). To our knowledge, this is the first experimental evidence of the wavelength-scale confinement of a surface wave propagating at the interface between two PTIs.

Finally, we demonstrate the topological protection of the surface wave by experimentally observing its most important physical property: that reflection-free energy flow can occur despite encountering a broad class of possible lattice defects along its propagation path that maintain spin-degeneracy [11, 12] and preserve the spin DOF (section S3 of the SI). Within this class falls the detour defect shown in the inset of Fig. 2b, where the rods are re-attached to the opposite plate so as to bend the interface between PTIs. The defect contains four 120° bends, each of which is capable of reflecting most of the incident surface wave in the absence of topological protection. As we show below, the addition of the defect creates a reflection-free single-channel delay line. The topological protection is apparent from Fig. 2b, where the transmission spectra along the uninterrupted interface (red line) and the same interface interrupted by a detour-type defect (green line) are plotted as a function of frequency. Outside of the bandgap (e.g., at the frequencies marked by black arrows) the transmission is reduced by almost an order of magnitude because the defect blocks the propagating bulk modes from the receiving antenna. Inside the bandgap, however, the forward-propagating spin-up TPSW flows around the defect (Fig.2c) because the defect does not flip the spin, and no back-reflection is allowed. The almost negligible $\sim 1\text{dB}$ decrease in transmission (red arrows) serves as a clear experimental signature of topologically robust transport.

The uniqueness of topological protection of guided waves is underscored by comparing the topological waveguide described above with a topologically-trivial one, which is formed by removing one row of cylinders. The resulting waveguide (orange box inset in Fig. 3b) supports topologically trivial guided waves (TTGWs) that are spectrally located (orange curve in Fig. 3a) inside the topologically trivial band gap of the claddings centered at $f_G^{(\text{tri})} \approx 3\text{GHz}$ (gray area). To emulate a time-delay line, we insert a large sharply-edged scattering defect (blue box inset of Fig. 3b) into the path of the TTGW. While the defect is almost identical to the one inserted into the path of the TPSW, its effect on the TTGW is dramatically different. For most frequencies, the measured transmission plotted in Fig. 3b (blue curve) is more than an order of magnitude lower than in the absence of the defect. High narrow-band transmission and tight spatial confinement (Fig. 3c) due to resonant tunneling [27, 28] and multi-bounce backscattering only occurs at the three Fabry-Pérot (FP) resonances of the defect. The absence of such FP resonances for the TPSWs observed from Fig. 2b is the experimental proof of their topological protection against backscattering.

These qualitative differences between topologically protected and trivial photon transport mechanisms motivates the usage of TPSW-based broadband delay lines. For example, a detour-type lattice defect shown in Fig. 2b introduces a time delay [29]

$$\tau = \frac{1}{2\pi}\left(\frac{\partial \varphi_{\text{def}}}{\partial f} - \frac{\partial \varphi_{\text{str}}}{\partial f}\right), \quad (1)$$

where $\varphi_{\text{str}}(f)$ and $\varphi_{\text{def}}(f)$ are the phases of the transmitted EM waves for the straight and defect-interrupted interfaces. The time delay can be indefinitely increased by stacking multiple detour defects (Fig.4). For example, two detours produce twice the delay time for the transmitted TPSWs, with no decrease in the bandwidth (Fig. 4a) because of the lack of interaction [22] between adjacent defects due to topological protection. On the contrary, using two detour-type defects in a topologically trivial waveguide reduces the operational bandwidth (Fig. 4b) for TTGWs.

The key advantage of topologically protected delay lines is their compactness: TPSWs can make sharp turns into tightly-packed phase-delaying detours without any backscattering. By loading a straight PTI interface (which plays the role of a bus waveguide [22] for the propagating TPSWs) with a sequence of phase-delay defects (Fig.4, top), nearly arbitrary phase profiles $\phi_{\text{def}}(x, f)$ can be generated and subsequently used in a variety of applications, including frequency-division multiplexing for free-space wireless communications [30] and terahertz wave generation [31]. The detour defect shown in Fig.2b,c can be viewed as a building block for these applications.

The experimentally measured phase and time delays of the transmitted TPSWs are plotted in Fig.5a for the straight and defect-interrupted interfaces. Note that $\tau(f) > 0$ everywhere inside the gap, although considerable fluctuations are observed at the edges of the gap, where the amplitude of the transmitted TPSW is reduced due to incomplete topological protection. However, for most frequencies the time delay is spectrally flat. In contrast, without topological protection, the delay time of TTGWs is a rapidly-varying sign-changing [24] function of the frequency inside the topologically trivial bandgap (Fig. 5b). These experimental results constitute the first step in building a multi-stage broadband topologically protected delay line capable of buffering multiple electromagnetic pulses.

In conclusion, we experimentally realized in the microwave frequency range a delay line based on a quantum spin Hall photonic topological insulator. Topological protection of localized surface waves between two PTIs resulted in reflectionless spectrally uniform time delays of several wave periods that were induced by a compact sharply-edged detour-type defect. Because such defects can be of nearly arbitrary shapes and sizes, we anticipate that novel geometries for compact wave buffers and delay lines utilizing topological photonic transport will emerge across the electromagnetic spectrum, from micro- to infrared waves. The introduced paradigm of an interface between two PTIs directed along a curved pathway will also enable creating near-arbitrary EM phase distributions $\phi(x,y,f)$ in the plane of a waveguide. By creating slit openings in the waveguide's wall, such phase distributions can be translated into complex frequency-dependent far-field patterns that can be utilized for frequency demultiplexing. Dynamic reconfiguration of the delay paths by rapid microelectromechanical displacement of the rods is also envisioned.

## Methods

### Transmission Measurements and Data Processing

Transmission measurement is performed with a linear dipole antenna as the feeding source which is inserted into the photonic structure through a small hole drilled through the top metallic cladding. When TPSWs are launched by such feeding source placed at the $x = x_0$ location, the two spin-up waves (with positive and negative refractive indexes) are launched in the forward direction toward the receiving probe. The receiving probe is also a linear dipole antenna which is placed on the outside at the end of the structure. We use a 2-port VNA (Agilent E5071C) to extract the amplitude and phase of the transmitted EM waves from the measured raw $|S_{12}|^2(y, f; x_0)$ spectra as described below. The spectra are smoothed with a 50 MHz window to suppress spectral contribution of VNA cables.

The subsequent data analysis aims at smoothing the interference pattern produced by the superposition of the positive and negative index spin-up TPSWs. The spatial periodicity of the resulting intensity pattern is $P_x = 3a$ along the interface separating the two PTIs. The following averaging treatment is used to suppress this beat pattern and to clarify the intensity distribution shown in Fig.2: every transmission curve of the TPSWs is the average of 6 individual raw $|S_{12}|^2(y, f; x_0)$ spectra for the feeding source positions $x_0^{(i)}$ (where $i = 1, \dots 6$) that are evenly spaced inside the $0 < x_0^{(i)} < P_x$ interval. In the case of TTGWs, similar procedure of averaging over the $0 < x_0^{(i)} < P_x$ interval is used, but with $i = 1,2,3$. The averaging domain spans from 4th rod to 6th rod along the interfaces for both TPSWs and TTGWs cases. The phases plotted in Fig.5 are extracted from the complex-valued $S_{12}(y, f; x_0^{(i)})$ parameters are taken directly from the raw spectra without any spatial averaging (i.e. $i = 1$). Note that the time delays for the TTGWs can be negative around resonances of the detour [24].

### Edge Scan Measurements

The entire fabricated structure is designed to be 45 periods along the x-direction for long range transport of the surface states, and 20 periods in the y-direction to prevent energy leakage to the lateral boundaries. The receiving probe is placed on the outside at the end of the structure, and can be scanned in the y-direction to map out the transverse profile of the transmitted waves for both the topologically non-trivial waveguide shown in Figs.1,2, and the topologically trivial waveguide shown in Fig.3. The receiving probe is mounted on a motorized stage (VELMEX Single Axis BiSlide) programmed together with the VNA to collect a single $S_{12}(y, f; x_0)$ spectrum, and then to reposition the probe by one step $\Delta y = 1.59$mm. Each complete scan consists of $N_y = 257$ spatial steps totaling $L_y = N\Delta y = 406.64$ mm span).

### Numerical Simulations

Two types of first-principles frequency domain electromagnetic simulations were performed using the COMSOL package: (a) the eigenvalue simulations which determine the frequencies $\omega(\vec{k}_\perp)$, ($\vec{k}_\perp = (k_x, k_y)$ is the in-plane wavenumber, $\omega = 2\pi f$ is the angular frequency), and (b) the driven simulations that determine the electric/magnetic field distribution for a given current source. To obtain the photonic band structures (PBS) of the topologically trivial/non-trivial waveguides, eigenvalue simulations were carried out on a supercell containing a single period along the x-direction, and 30 cells on each side of the interface. The shaded regions in Fig.1b and Fig.3c are the projected band structure of the bulk modes. The transmission through the photonic structures containing a straight interface and a bent interface with one or two detours were calculated using driven simulations on the simulation domains containing 20x45 cells so as to closely approximate the actual structure used in the experiments.

To examine the effect of Ohmic loss in metal and plasmonic effects on the propagation of the TPSWs around $\lambda = 1.5\mu m$, we performed eigenfrequency simulation by scaling down the dimensions of the photonic structure. The relative geometric sizes of the structure were slightly adjusted to account for the plasmonic response of the selected metal (silver). With loss in the metal fully accounted for, we found the propagation length $L_x \approx 60a \approx 48\lambda$ to be sufficiently long for taking advantage of topological protection.

**Reference**

1. Haldane, F. & Raghu, S., Possible realization of directional optical waveguides in photonic crystals with broken time-reversal symmetry. *Phys. Rev. Lett.* **100**, 013904 (2008).

2. Poo, Y., Wu, R., Lin, Z., Yang, Y. & Chan, C. T., Experimental Realization of Self-Guiding Unidirectional Electromagnetic Edge States. *Phys. Rev. Lett.* **106**, 093903 (2011).

3. Raghu, S. & Haldane, F., Analogs of quantum-Hall-effect edge states in photonic crystals. *Physical Review A* **78** (3), 033834 (2008).

4. Wang, Z., Chong, Y., Joannopoulos, J. & Soljačić, M., Reflection-free one-way edge modes in a gyromagnetic photonic crystal. *Phys. Rev. Lett.* **100**, 13905 (2008).

5. Wang, Z., Chong, Y., Joannopoulos, J. D. & Soljačić, M., Observation of unidirectional backscattering-immune topological electromagnetic states. *Nature* **461**, 772-775 (2009).

6. Hafezi, M., Demler, E. A., Lukin, M. D. & Taylor, J. M., Robust optical delay lines with topological protection. *Nature Physics* **7** (11), 907-912 (2011).

7. Hafezi, M., Mittal, S., Fan, J., Migdall, A. & Taylor, J. M., Imaging topological edge states in silicon photonics. *Nature Photonics* **7**, 1001–1005 (2013).

8. Umucalılar, R. & Carusotto, I., Artificial gauge field for photons in coupled cavity arrays. *Physical Review A* **84** (4), 043804 (2011).

9. Fang, K., Yu, Z. & Fan, S., Realizing effective magnetic field for photons by controlling the phase of dynamic modulation. *Nature Photonics* **6** (11), 782-787 (2012).

10. Rechtsman, M. C. *et al.*, Photonic Floquet topological insulators. *Nature* **496** (7444), 196-200 (2013).

11. Khanikaev, A. B. *et al.*, Photonic topological insulators. *Nature Materials* **12** (3), 233-239 (2013).

12. Chen, W.-J. *et al.*, Experimental realization of photonic topological insulator in a uniaxial metacrystal waveguide. *Nature Communications* **5** (2014).

13. Lu, L., Joannopoulos, J. & Soljačić, M., Topological photonics. *Nature Photonics* **8**, 821 (2014).

14. Ma, T., Khanikaev, A. B., Mousavi, S. H. & Shvets, G., Guiding Electromagnetic Waves around Sharp Corners: Topologically Protected Photonic Transport in Metawaveguides. *Phys. Rev. Lett.* **114** (12), 127401 (2015).

15. Kane, C. L. & Mele, E. J., Quantum Spin Hall Effect in Graphene. *Phys. Rev. Lett.* **95**, 226801 (2005).

16. Bernevig, B. A. & Zhang, S.-C., Quantum Spin Hall Effect. *Phys. Rev. Lett.* **96**, 106802 (2006).

17. Hsieh, D. *et al.*, A topological Dirac insulator in a quantum spin Hall phase. *Nature* **452** (7190), 970-974 (2008).

18. König, M. *et al.*, Quantum spin Hall insulator state in HgTe quantum wells. *Science* **318** (5851), 766-770 (2007).

19. Wang, Z. & Fan, S., Optical circulators in two-dimensional magneto-optical photonic crystals. *Opt. Lett.* **30**, 1989 (2005).

20. Qiu, W., Wang, Z. & Soljačić, M., Broadband circulators based on directional coupling of one-way waveguides. *Opt. Express* **19**, 22248 (2011).

21. Wallace, J. W. & Jensen, M. A., Mutual Coupling in MIMO Wireless Systems: A Rigorous Network Theory Analysis. *IEEE TRANSACTIONS ON WIRELESS COMMUNICATIONS* **3**, 1317 (2004).

22. Xia, F., Sekaric, L. & Vlasov, Y., Ultacompact optical buffers on a silicon chip. *Nature Photonics* **1**, 65 (2007).

23. Bogaerts, W. *et al.*, Silicon microring resonators. *Laser Photonics Rev.* **1**, 47 (2012).

24. Mittal, S. *et al.*, Topologically Robust Transport of Photons in a Synthetic Gauge Field. *Phys. Rev. Lett.* **113**, 087403 (2014).

25. Chabanov, A. & Genack, A., Statistics of Dynamics of Localized Waves. *Phys. Rev. Lett.* **87**, 233903 (2001).

26. Sheng, D. N., Weng, Z. Y., Sheng, L. & Haldane, F. D. M., Quantum Spin-Hall Effect and Topologically Invariant Chern Numbers. *Phys. Rev. Lett.* **97** (3), 036808 (2006).

27. Fan, S., Villeneuve, P. R., Joannopoulos, J. D. & Haus, H., Channel drop tunneling through localized states. *Physical Review Letters* **80** (5), 960 (1998).

28. Meade, R. D., Brommer, K. D., Rappe, A. M. & Joannopoulos, J., Electromagnetic Bloch waves at the surface of a photonic crystal. *Physical Review B* **44** (19), 10961 (1991).

29. Sebbah, P., Legrand, O. & Genack, A., Fluctuations in photon local delay time and their relation to phase spectra in random media. *Phys. Rev. E* **59**, 2406 (1999).

30. Karl, N. J., McKinney, R. W., Monnai, Y., Mendis, R. & Mittleman, D. M., Frequency-division multiplexing in the terahertz range using a leaky-wave antenna. *Nature Photonics* **9**, 717 (2015).

31. Bhattacharjee, S. *et al.*, Folded Waveguide Traveling Wave Tube Sources For Terahertz Radiation. *IEEE Transactions On Plasma Science* **32**, 1002 (2004).

**Figure**

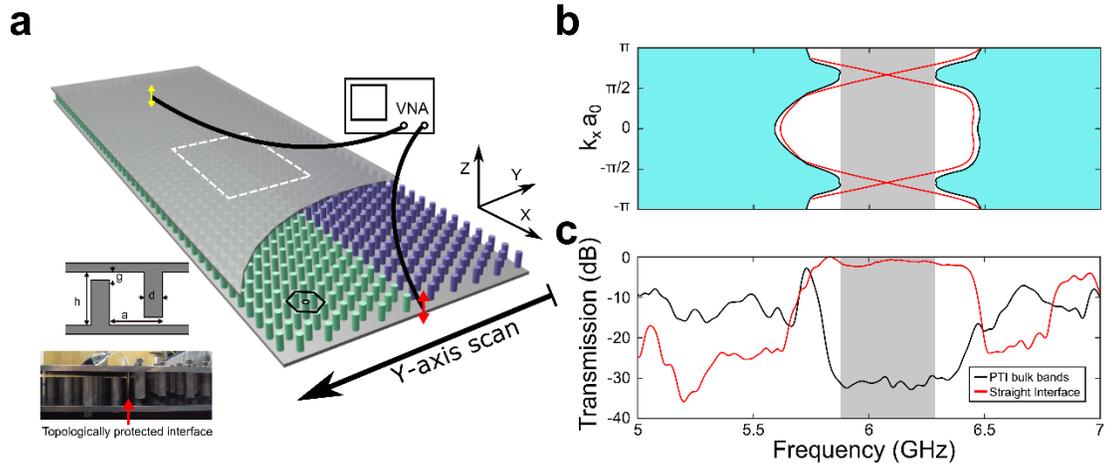

**Figure 1| The platform for a quantum spin-Hall photonic topological insulator (QSH PTI): a bianisotropic metawaveguide.** **(a)** Schematic of a QSH PTI. Part of the top metal plate is removed to reveal the cylinders attached to the top plate (purple) and to the bottom plate (green), leaving a gap of thickness $g$ to the other plate. The 2-port VNA is connected to the feeding source (yellow double arrow) and to the receiving probe (red double arrow) for transmission measurement. Top inset: geometric parameters of the PTI. Bottom inset: picture of the assembled structure showing an interface between the two the PTIs which serve as two topological claddings. **(b)** Calculated 1D projected PBS of the PTI with topologically non-trivial interface. Cyan area: bulk bands, gray area: a complete band gap around the doubly degenerate Dirac cones, red curve: TPSWs supported by the topologically non-trivial interface. **(c)** Measured transmission spectra through the bulk PTI (all rods attached to the top plate: black curve) and along the interface between two PTIs shown in Fig. 1a (red curve). The transmission is enhanced by nearly 30 dB in the $5.87 < f < 6.29$ GHz frequency range by the presence of the interface, indicating that the surface propagation dominates over the bulk PTI propagation. QSH PTI parameters defined in the inset: $h = a = 36.8$ mm, $d = 0.345\,a = 12.7$ mm, $g = 0.15\,a = 5.5$ mm.

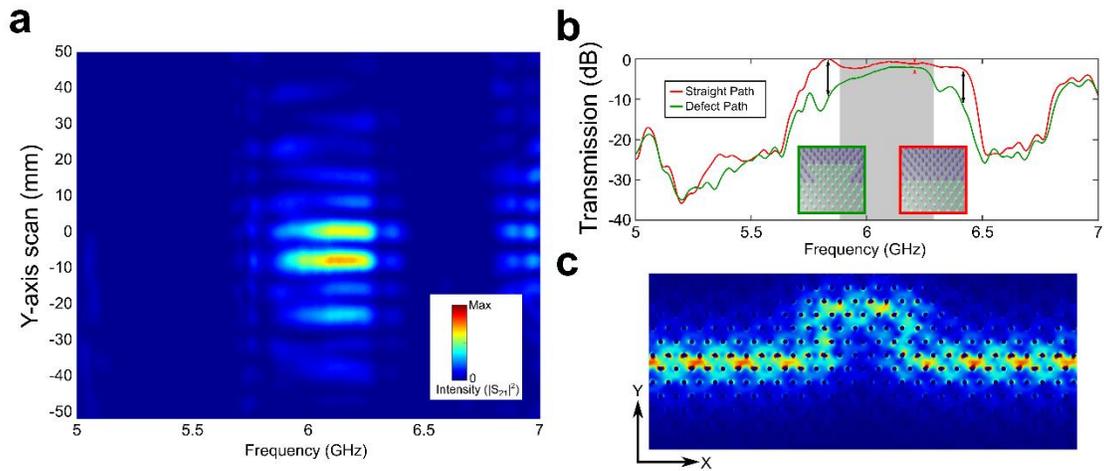

**Figure 2| Spatial localization and topological protection of the surface waves. (a)** Frequency-dependent spatial profiles of the electric field intensity measured at the end of the photonic structure. The surface mode is transversely confined to a small fraction of its wavelength. The zero of the scanning axis ($y = 0$) is at the interface between the two PTIs. **(b)** Comparison of measured transmission spectra between the straight (red curve) and the delay line (green curve) interfaces. The receiving probe is at $y = 0$. Inset: schematic of the interface inside the dashed white box in Fig. 1a for the straight interface (red box) and the delay line interface (green box) containing a large detour-type defect. **(c)** Simulated energy density at $f = 6.08$ GHz showing the TPSW flowing around the defect without scattering.

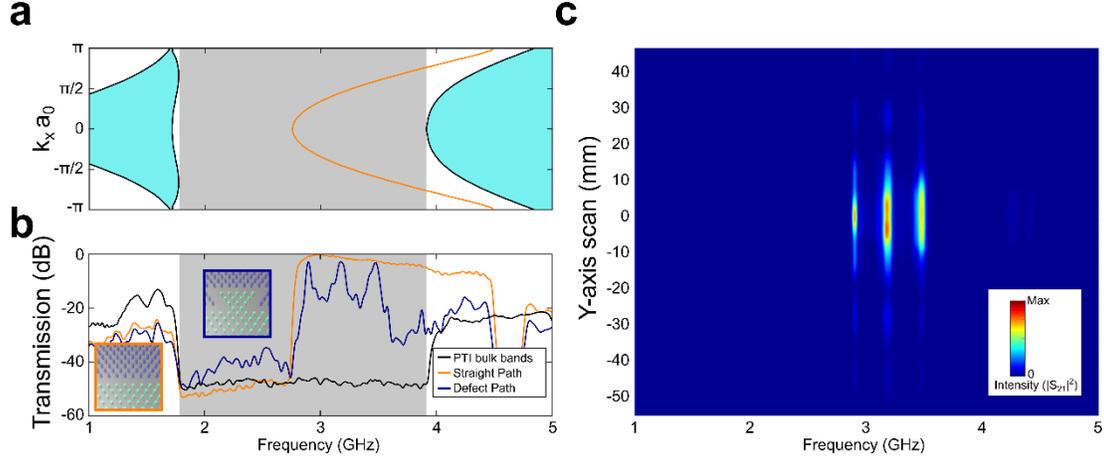

**Figure 3. Properties of the guided modes of a topologically trivial waveguide produced by removing one row of rods.** The waveguide shown in the orange box in the inset of (b) is classified as topologically trivial because the bulk modes of the *claddings* possess a vanishing Chern index. The center frequency $f_G^{(\text{tri})} \approx 3\text{GHz}$ of the topologically trivial bandgap is considerably lower than that of the topologically nontrivial bandgap $f_G^{(\text{top})} \approx 6.1\text{GHz}$ shown in Fig.1. **(a)** Calculated 1D projected PBS: bulk bands of the claddings (cyan area), a complete band gap (gray area), and the dispersion curve of the TTGW (orange lines). **(b)** Orange box inset: straight waveguide, blue box inset: bent waveguide. Black curve: transmission spectrum through the bulk of the *claddings* (i.e. a row of rods is not removed). Transmission drop in the $1.8 < f < 3.9\text{GHz}$ range indicates the complete bandgap. Near-perfect spectrally-flat transmission (orange curve) through the straight waveguide indicates reveals TTGWs in the $3.0 < f < 3.9\text{GHz}$ range. Transmission through the bent waveguide (blue curve) is negligible for all frequencies except at $f_1^{(\text{FP})} = 2.9\text{GHz}$, $f_2^{(\text{FP})} = 3.2\,\text{GHz}$, and $f_3^{(\text{FP})} = 3.4\,\text{GHz}$ corresponding to Fabry-Perot resonances of the defect. High transmission at $f_{1,2,3}^{(\text{FP})}$ is enabled by multiple bounces of the TTGW inside the defect. **(c)** Frequency-dependent spatial profiles of the electric field intensity measured at the end of the photonic structure. Spatially confined guided modes are observed in three narrow frequency bands.

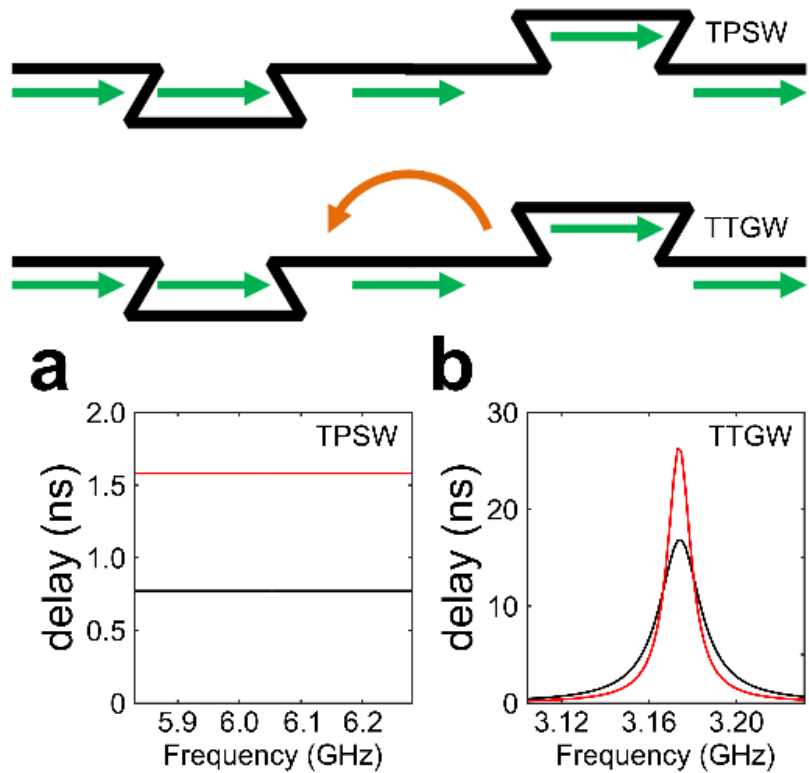

**Figure 4 | Photonic time delay lines based on detour-type defects**. **Top:** schematics of stacking two defects along the photonic waveguides. The second defect is placed at a distance $\Delta x = 16a$ after the first defect. Green arrows: transmitted, orange arrows: reflected EM waves. **(a)** Two compact detours (red line) produce twice the delay time of the one (black line), with no decrease in the bandwidth for TPSWs. **(b)** In contrast, two detour-type defects in a topologically trivial waveguide reduces the operational bandwidth because of the reflections of TTGWs. While the peak delay is almost doubled by the second defect, the bandwidth is reduced to less than half of the original value. The resultant delay-bandwidth product $N = \Delta f \times \tau$, which determines the number of electromagnetic pulses that can be buffered by a delay line, does not increase by the addition of the second defect.

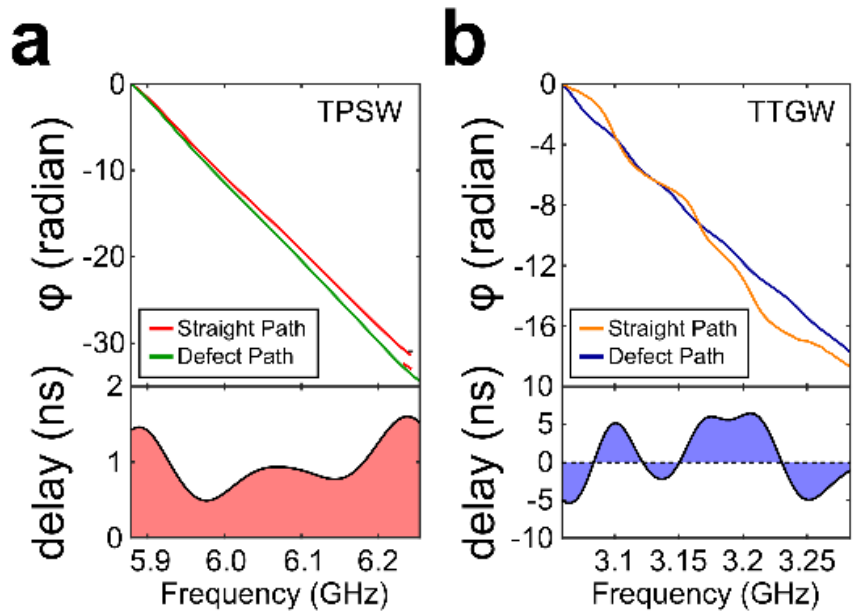

**Figure 5. Photonic delay line applications based on QSH PTI.** Top half: the relative phases extracted from the transmission spectra of straight path and defect; bottom half: the time delays of delay line application on the platforms of **(a)** TPSWs and **(b)** TTGWs. Note that the time delays for the TTGWs can be negative around the Fabry-Perot (FP) resonances of the detour [24]. Negative time delays are not observed for the detoured TPSWs because topological protection against backscattering prevents FP resonances.

# Supplementary Information

# Experimental Realization of a Reflections-Free Compact Delay Line Based on a Photonic Topological Insulator

Kueifu Lai[1], Tzuhsuang Ma[1], Xiao Bo[2,3], Steven Anlage[2,3], and Gennady Shvets[1,*]

[1] Department of Physics, The University of Texas at Austin, Austin, Texas 78712, USA

[2] Center for Nanophysics and Advanced Materials, Department of Physics, University of Maryland, College Park, Maryland 20742-4111, USA

[3] Department of Electrical and Computer Engineering, University of Maryland, College Park, Maryland 20742-3285, USA

# 1. Design and properties of the Quantum Spin Hall Photonic Topological Insulator

The photonic topological insulator (PTI) used in this work is based on the bianisotroic meta-waveguide (BMW) platform [1]. It consists of a hexagonal array of metal rods sandwiched between two parallel metal plates separated by the distance $h_0$ that confine the electromagnetic waves in the vertical (z-) dimension. When the posts are symmetrically placed between two metal plates (and separated from them by the gap $g_0$ as shown in Fig.S1(a)), the structure can be viewed as "photonic graphene" (PhG) [2] with the period $a_0$ [3]. The PhG structure supports transverse electric (TE) and magnetic (TM) modes shown in Fig.S1(b). The modes differ from each other by their symmetry properties with respect to mid-plane reflection operation $\sigma_z$: the $H_z$ component of the TE (TM) mode is symmetric (anti-symmetric) with respect to $\sigma_z$ while the opposite holds for the $E_z$ components of the two modes.

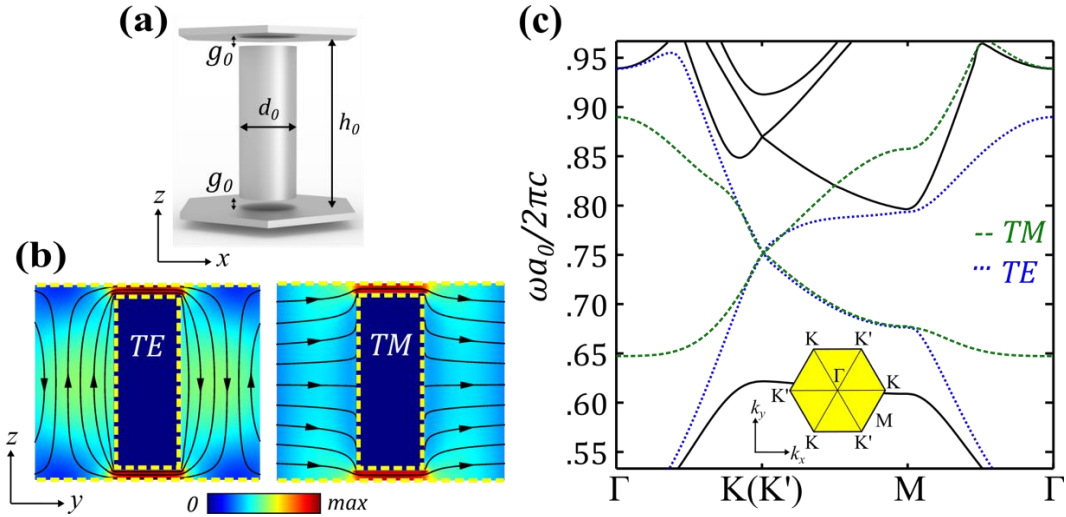

**Figure S1:** The unperturbed "photonic graphene" (PhG) structure used for emulating photon equivalents of the spin and valley degrees of freedom. (**a**) The unit cell of the PhG: metal rods arranged as a hexagonal array lattice with the lattice constant $a_0$. (**b**) Magnetic field profiles of the TE and TM modes at the $K$ point. (**c**) The PBS with TE and TM modes forming doubly-degenerate Dirac cones at $K(K')$ points. Design parameters: $h_0 = a_0$, $d_0 = 0.345 a_0$, and $g_0 = 0.05 a_0$.

Each of the two modes is doubly-degenerate for $\mathbf{k}_\perp = \pm \mathbf{e}_x 4\pi/3a_0$ corresponding to the $K(K')$ edges of the Brillouin zone shown in the inset to Fig.1S(b). The hexagonal symmetry of the PhG lattice guarantees the appearance of the Dirac cone for the decoupled TE/TM modes. Moreover, for a given period $a_0$, the two modes can be made degenerate with each other at the $K(K')$ points by the judicious choice of $h_0$ and the cylinders' diameter $d_0$. Such mode-degeneracy is essential [4] for establishing spin-like linear combinations of the TE/TM modes which can be coupled to each other by a bi-anisotropic perturbation of the photonic structure. One such perturbation involves simply closing one of the gaps between the rods and the plates, i.e. attaching the rod to one of the plates as shown in Fig.S2. The resulting band gap shown in Fig.S2(b) separates the photonic phases with the opposite signs of the spin-Chern number $C^{SOC}_{\uparrow/\downarrow} = \pm 1 \times \text{sgn}(\Delta_{SOC})$, where is the bianisotropic coefficient whose sign is determined by the plate to which the metal rod is attached [1]: $\Delta_{SOC} > 0$ if the rod is attached to the bottom plate, and $\Delta_{SOC} < 0$ if the rod is attached to the top plate. The size of the bandgap is also proportional to $|\Delta_{SOC}|$. For those frequencies inside the bandgap, this photonic structure behaves as a photonic topological insulator emulating the quantum spin Hall effect (QSH-PTI). The bianisotropy parameter $\Delta_{SOC}$ in PTIs emulates the strength of the spin-orbit interaction in the Kane-Mele model [5] of spin-Hall effect in graphene, which controls the Chern number of the topological phases.

Note that the spin-Chern number is not a global topological index because the sum of the spin-Chern numbers over all spin states vanishes: $\sum_{k=\downarrow,\uparrow} C^{SOC}_k = 0$. However, for the spin-degenerate bianisotropic structures [1] considered here there is no inter-spin scattering. Therefore, two independent copies of the topological phases with opposite spin-Chern numbers can be independently considered.

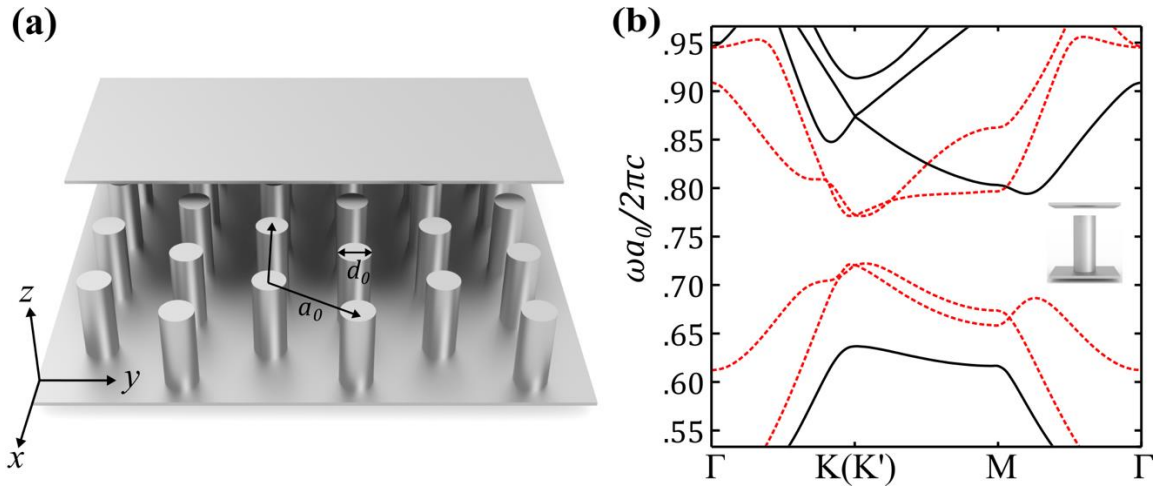

**Figure S2: Bi-anisotropic meta-waveguide (BMW) as a photonic topological insulator.** **(a)** Schematic of the BMW. Part of the top metal plate is removed to reveal the "bed-of-nails" structure below. **(b)** PBS with the bandgap induced by the bianisotropy of the meta-waveguide: gap between the rod and the bottom metal plate is closed. Dashed lines: the hybridized TE/TM bands of interest. BMW parameters: $h_0 = a_0$, $d_0 = 0.345 a_0$, $g_0 = 0.15 a_0$.

The ability to engineer the sign of the Chern number by making a simple change to the structure (e.g., by reversing the orientation of its constituent elements [4] or by attaching the rods to a different plate [1]) is a unique feature of bianisotropic PTIs that are not available with many other designs [6] where the sign of the effective spin-orbit interaction is fixed.

## 2. Emergence of topologically protected surface waves at the domain wall between two QSH-PTIs

The existence and number of the topologically protected surface waves (TPSWs) at a QSH/QSH interface can be predicted based on the bulk-boundary correspondence principle [ 7]. One such interface, between a QSH with $\Delta_{SOC}^{(1)} > 0$ and a QSH with $\Delta_{SOC}^{(2)} < 0$, is shown in Fig.S3(a) a thin dashed line. All together there two TPEWs moving in a given direction because $\Delta C = C_{\uparrow(\downarrow)}^{(1)} - C_{\uparrow(\downarrow)}^{(2)} = \pm 2$. Therefore, spin-up TPSWs are moving to the right and spins-down TPSWs are moving to the left along the interface as predicted by the corresponding photonic band structure shown in Fig.S3(b). The field profiles of the four TPSWs labeled by their spins and refractive indices are shown in Fig.S3(c,d) [ 3].

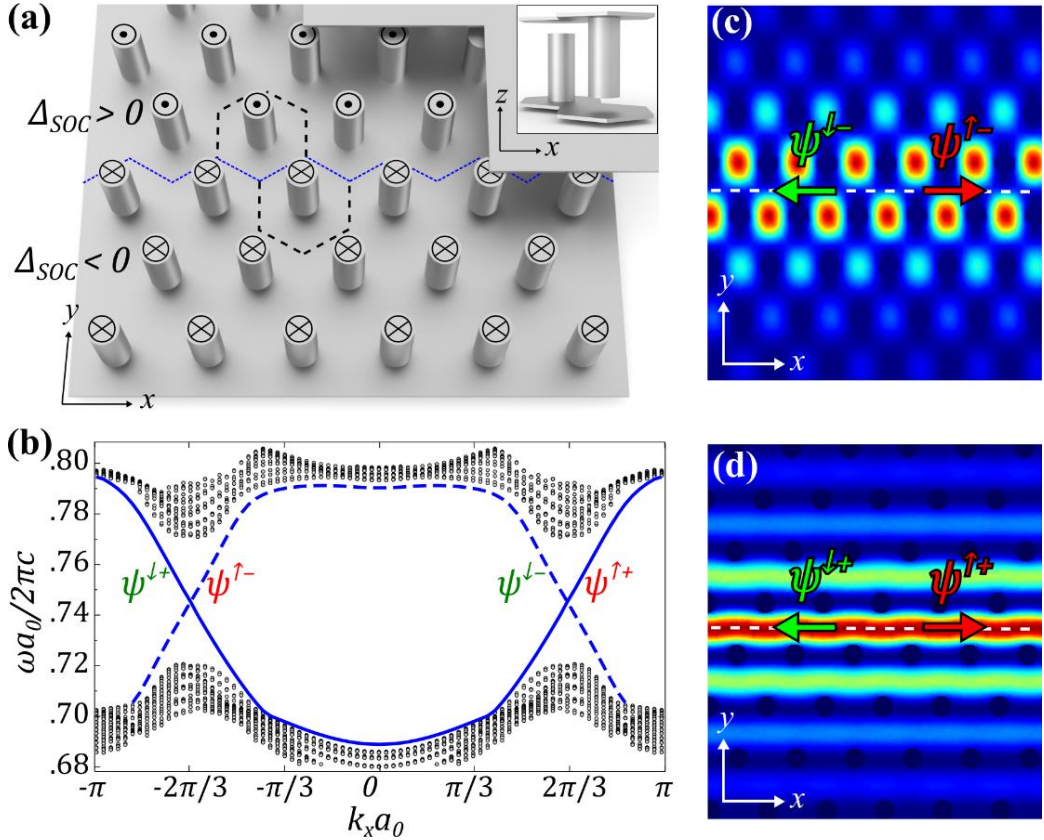

**Figure S3: Topologically protected edge modes along QSH/QSH zigzag interface.** (a) Top views of the zigzag interface (thin dashed line) between two QSH PTIs investigated in this Letter. Thick dashed line outlines the side view (inset) of two adjacent unit cells on the edge. Black arrows on rods indicate $\text{sgn}(\Delta_{SOC})$. (b) PBS of a supercell (single cell along *x*-direction, 30 cells on each side of the interface). Black circles: bulk modes, blue solid/dashed lines: dispersion curves of the TPSWs with positive/negative refractive index. ↑/↓ labels mark the spin, and +/− labels mark the sign of the refractive index $n(\omega) \equiv ck_x/\omega$. (c,d) Field profiles of the TPEWs with positive/negative refractive index respectively. Color: $|E_z \pm H_z|^2$ for ↑/↓ respectively. Red and green arrows show the propagation direction of the edge mode with different spins.

It can be observed from Figs.S3(c,d) that the propagation direction of TPSWs is spin-locked. The two types of spin-up forward-propagating TPSWs have the same group velocity but different phase velocities: one has a positive refractive index and another one has a negative one. In the experiment we are simultaneously exciting both forward-propagating TPSWs because we are using a dipole antenna for excitation. A more sophisticated phased antennas array could be used in the future for exciting just one of the two modes. In addition, one could use a far-field excitation of TPSWs using a narrow slit cut through the top plate as explained in Ref. [ 8]. An electromagnetic wave incident on the slit at an angle $\phi(\omega)$ with respect to the normal (z-) direction that satisfies the phase matching condition with a given TPSW (i.e. $\sin \phi (\omega) = n(\omega)$ for a TPSW with the refractive index $n(\omega) = ck(\omega)/\omega$ ) would strongly couple to that particular TPSW. These two approaches to selective excitation of TPSWs is the subject of our future work.

## 3. Preservation of spin-degeneracy and spin conservation in the presence of defects

Because the spin degree of freedom in QSH-PTIs is synthetic, the topological protection of TPSWs does not directly follow from time-reversal symmetry as it does in conventional topological insulators. Instead, TPSWs rely on the properties of spin-degeneracy and the conservation of the spin DOF [ 1, 4]. Only a limited class of defects ensure these properties. One such lattice defect involves random variations of the magnitude of the effective spin-orbit coupling coefficient $\Delta_{SOC}$, which itself is determined by the gap size $g$ between the metal rod and plate. Changing the size of the gap can be viewed as an electromagnetic perturbation that affects the TE and TM modes differently, thus potentially violating spin-degeneracy. Our earlier analytic estimates quantified [ 1] the strength of thus induced spin coupling and compared it to the strength of the effective spin-orbit coupling $\Delta_{SOC}$. The former was predicted to be much smaller than the latter.

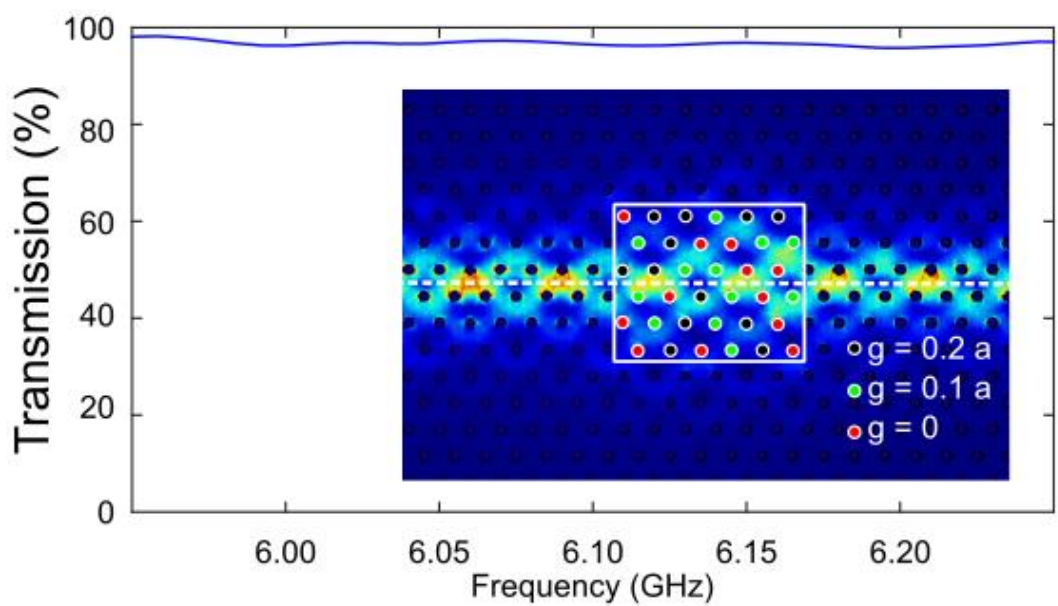

**Figure S4| Performance of QSH PTI with Random Gap Sizes.** Calculated transmission spectrum of TPSWs propagating along the straight interface through a region of defects with random gap size: $g \neq 0.15a$. Blue Curve: transmission of TPSWs. The transmission is large than 90% over the frequency range of interest. Inset: Energy density of TPSWs propagation. White dashed line: the domain wall between two QSH PTIs (top domain: rods attached to top plate, bottom domain: rod attached to bottom plate). White-bordered box: defect region with color-coded random gap sizes (black: $g = 0.2a$, green: $g = 0.1a$, red: $g = 0$). The TPSWs are excited by a point dipole to the left of the shown domain. Parameters of the QSH PTIs are the same as in the caption to Fig.1 everywhere, except inside the white-bordered box.

To investigate the implications of this near-conservation of the spin DOF, we have devised a numerical experiment shown in Fig.S4. The numerical experiment uses the same sizes (listed in the caption of Fig.1 of the Letter) as the experiment, except that a very large portion of the photonic lattice is perturbed. The perturbed region inside the white-bordered box in Fig.S4 consists of 35 rods ($\approx 4\lambda \times 4\lambda$) near the domain wall between the two PTIs with opposite values of $\Delta_{SOC}$ has been subjected to strong and random perturbation of the gap size $g$: the baseline value $g = 0.15a$, the varied gap is $0 < g < 0.2a$. Within our computational accuracy, such strong defect does not reduce transmission and does not cause any detectable back-scattering of the TPSWs. That implies that there is no mixing of the spin states by the perturbation and, therefore, no violation of the topological protection. Note that this perturbation does not involve any displacement of the rods from their positions on the hexagonal lattice.

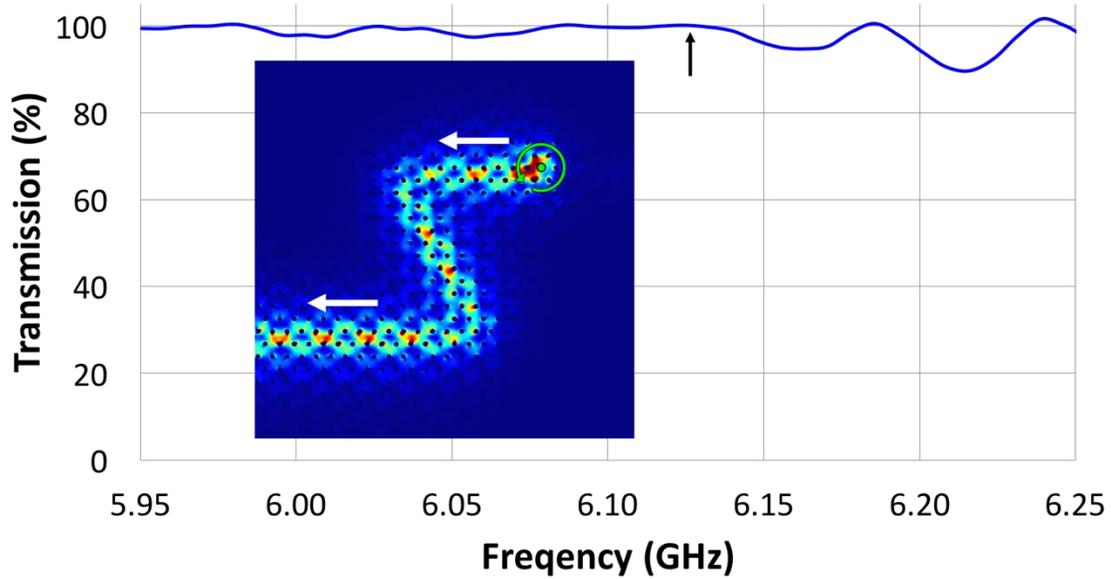

**Figure S5: Propagation of TPSWs along the domain wall between two QSH PTIs with opposite signs of the effective spin-orbit coupling.** Transmission spectrum $T(\Delta\omega)$ through the zigzag-type path, where $\Delta\omega = \omega - \omega_G$ is the detuning from the bandgap center at $\omega_G a_0/2\pi c = 0.745$. The spin-down TPSWs are excited by placing an electric dipole between the rod and metal plate in the upper right corner. Parameters of the QSH PTIs are the same as in the caption to Fig.1.

The second defect type is the one which is the subject of this Letter: a deformation of the domain wall between the two PTIs. That is, one can design an arbitrarily-shaped domain wall with multiple sharp bends of the propagation path, and the guided surface wave will still propagate through without any backscattering. One such defect, which is a combination of multiple sharp turns of the domain wall, has been theoretically for the case of the two $120°$ zigzag-type bends analyzed [ 1]. These simulations results are reproduced in Fig.S5.

# References

1. Ma, T., Khanikaev, A. B., Mousavi, S. H. & Shvets, G., Guiding Electromagnetic Waves around Sharp Corners: Topologically Protected Photonic Transport in Metawaveguides. *Phys. Rev. Lett.* **114** (12), 127401 (2015).

2. Szameit, A., Rechtsman, M. C., Bahat-Treidel, O. & Segev, M., PT-symmetry in honeycomb photonic lattices. *Phys. Rev. A* **84**, 021806(R) (2011).

3. Ma, T. & Shvets, G., Scattering-Free Optical Edge States between Heterogeneous Photonic Topological Insulators. *arXiv preprint arXiv:1507.05256* (2015).

4. Khanikaev, A. B., Mousavi, S. H., Tse, W.-K., Kargarian, M., MacDonald, A. H. & Shvets, G., Photonic topological insulators. *Nature Materials* **12** (3), 233-239 (2013).

5. Kane, C. L. & Mele, E. J., Quantum Spin Hall Effect in Graphene. *Phys. Rev. Lett.* **95**, 226801 (2005).

6. Chen, W.-J., Jiang, S.-J., Chen, X.-D., Zhu, B., Zhou, L., Dong, J.-W. & Chan, C. T., Experimental realization of photonic topological insulator in a uniaxial metacrystal waveguide. *Nature Communications* **5** (2014).

7. Mong, R. S. & Shivamoggi, V., Edge states and the bulk-boundary correspondence in Dirac Hamiltonians. *Physical Review B* **83** (12), 125109 (2011).

8. Karl, N. J., McKinney, R. W., Monnai, Y., Mendis, R. & Mittleman, D. M., Frequency-division multiplexing in the terahertz range using a leaky-wave antenna. *Nature Photonics* **9**, 717 (2015).